\documentclass[iop,apj]{emulateapj}
\usepackage{amsmath}
\usepackage{ulem}

\usepackage[colorlinks=true,citecolor=blue,filecolor=black,runcolor=blue,linkcolor=blue,urlcolor=blue]{hyperref}

\shorttitle{The fate of a red nugget}
\lefthead{Morishita \& Ichikawa}

\begin{document}
\title{The fate of a red nugget: In-situ star formation of satellites around a massive compact galaxy}
\author{Takahiro Morishita\altaffilmark{1,2,3} and
	Takashi Ichikawa\altaffilmark{1}
	}
\email{mtakahiro@astr.tohoku.ac.jp}
\altaffiltext{1}{Astronomical Institute, Tohoku University, Aramaki, Aoba, Sendai 980-8578, Japan}
\altaffiltext{2}{Institute for International Advanced Research and Education, Tohoku University, Aramaki, Aoba, Sendai 980-8578, Japan}
\altaffiltext{3}{Department of Physics and Astronomy, University of California, Los Angeles, CA, 90095-1547, USA}

\begin{abstract}
To study the accretion phase for local massive galaxies, we search accreting satellites around a massive compact galaxy ($M_*\sim3.9\times10^{10}M_\odot$), spectroscopically confirmed ($z_{\rm spec}=1.9213$) in the eXtreme Deep Field, which has been originally reported in Szomoru~{et al.}
We detect 1369 satellite candidates within the projected virial radius ($r_{\rm vir}\sim300$~kpc) of the compact galaxy in the all-combined ACS image with $5\sigma$-limiting magnitude of $m_{\rm ACS}\sim30.6$~ABmag, which corresponds to $\sim1.6\times10^7 M_\odot$ at the redshift.
The photometric redshift measured with 12~multi-band images confirms 34~satellites out of the candidates.
Most of the satellites are found to have the rest-frame colors consistent with star forming galaxies.
We investigate the relation between stellar mass and star formation rate (the star formation main sequence), and find the steeper slope at the low-mass end ($<10^8M_\odot$), while more massive satellites are consistently on the sequence reported in previous studies.
Within the uncertainties of star formation and photometric redshift, we conjecture possible scenarios for the compact galaxy which evolves to a local massive galaxy by way of significant size and mass growth.
While merging of the existing total stellar mass of the satellites is not enough to explain the mass growth predicted by observations and simulations, the contribution by in-situ star formation in the satellites would compensate the deficit.
Provided that most satellites keep the observed in-situ star formation and then quench before they accrete by, e.g., environmental quenching, the compact galaxy would become a massive early-type galaxy consistent with the local size-mass relation.
\end{abstract}

\keywords{galaxies: high-redshift --- galaxies: evolution --- 
	galaxies: formation --- galaxies: structure}


\section{INTRODUCTION}
The formation scenario of local massive early-type galaxies (ETGs) has been pursued for half a century.
Major merger of massive galaxies has been the standard formation scenario, but only in highly dense regions, such as in clusters and groups (Dressler~\citeyear{dressler80}).
Recent understanding of local massive ETGs, including in the less dense field, is boosted by the studies of high redshift ($z$) massive galaxies.
A candidate for the progenitor of local massive ETGs is a compact galaxy, which is also known as ``red nugget'' (Damjanov {et al.}~\citeyear{damjanov09}).
The galaxy is characterized to have massive ($M_*\ga3\times10^{10}M_\odot$) and passively evolving old ($\ga0.5$~Gyr) stellar component with small scale radius ($r_e\sim1$~kpc).
The observational fact that such compact galaxies formed in the early universe, typically at $z>1.5$, has given us an important clue to the formation scenario of local massive ETGs.

Since compact galaxies have been reported to be rare in the local universe (e.g., Taylor~{et al.}~\citeyear{taylor10}), many studies have focused on explaining the evolution by linking to local massive ETGs.
However, the observed significant size growth by a factor of 2-5 (Trujillo~{et al.}~\citeyear{trujillo07}; Buitrago~{et al.}~\citeyear{buitrago08}; Hopkins~{et al.}~\citeyear{hopkins09b}; Patel {et al.}~\citeyear{patel13}) remains unexplained, in spite of much effort of theoretical studies (e.g., Fan {et al.}~\citeyear{fan08}), though being still under debate with, e.g., progenitor bias (Carollo {et al.}~\citeyear{carollo13}).
The significant evolution of compact galaxies has led to the idea on the formation scenario of local massive ETGs in ``two-phase" scheme, where massive stellar components, i.e. red nuggets, were formed first, followed by numerous accretion of gas poor (dry) stellar system to form outer envelope and enlarge the galaxy scale radius (Oser {et al.}~\citeyear{oser10},~\citeyear{oser12}; Toft {et al.}~\citeyear{toft14}).
The physical aspect of the former phase, or ``collapse phase", which makes up the compact galaxies in a short time duration ($\sim0.5$~Gyr; van Dokkum {et al.}~\citeyear{vandokkum10}), has been investigated with many observations (Barro {et al.}~\citeyear{barro13}; Tadaki {et al.}~\citeyear{tadaki14}; van Dokkum~{et al.}~\citeyear{vandokkum15b}).
The collapse phase is also predected by several numerical simulations (Dekel {et al.}~\citeyear{dekel14}; Wellons {et al.}~\citeyear{wellons15}), in concordance with the bulge formation (e.g., Noguchi~\citeyear{noguchi99}).
The following phase, or ``accretion phase", consistently explain the significant size evolution, in contrast to major merger which does not bring the local size-mass relation (Trujillo {et al.}~\citeyear{trujillo07}; Damjanov {et al.}~\citeyear{damjanov09}) and less frequently happens (Bundy~{et al.}~\citeyear{bundy09}; Lotz~{et al.}~\citeyear{lotz11}).
Several observations have challenged to find the evidence of the latter phase by directly searching accreting satellites within $\sim100$~kpc (or within virial radius, $r_{\rm vir}$) of massive galaxies, though finding the number deficit to explain the significant growth ({M{\'a}rmol-Queralt{\'o}} {et al.}~\citeyear{marmol12}; Bluck~{et al.}~\citeyear{bluck12}; Newman~{et al.}~\citeyear{newman12}).
Detecting less massive satellites ($M_*<10^{10} M_\odot$) around massive galaxies strongly depends on the observation depth and spatial resolution, while the robust measurement of the spectroscopic redshift of the companion would be further difficult.

\begin{figure*}
\figurenum{1}
\begin{center}
\includegraphics[width=7cm,bb= 0 0 240 228]{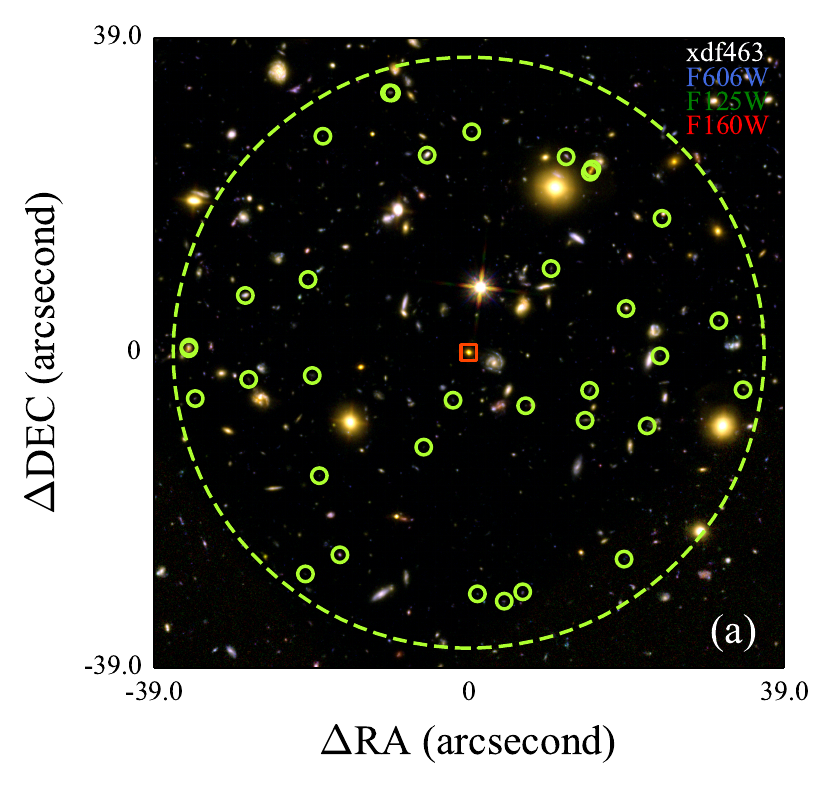}
\includegraphics[width=9cm,bb= 0 0 293 212]{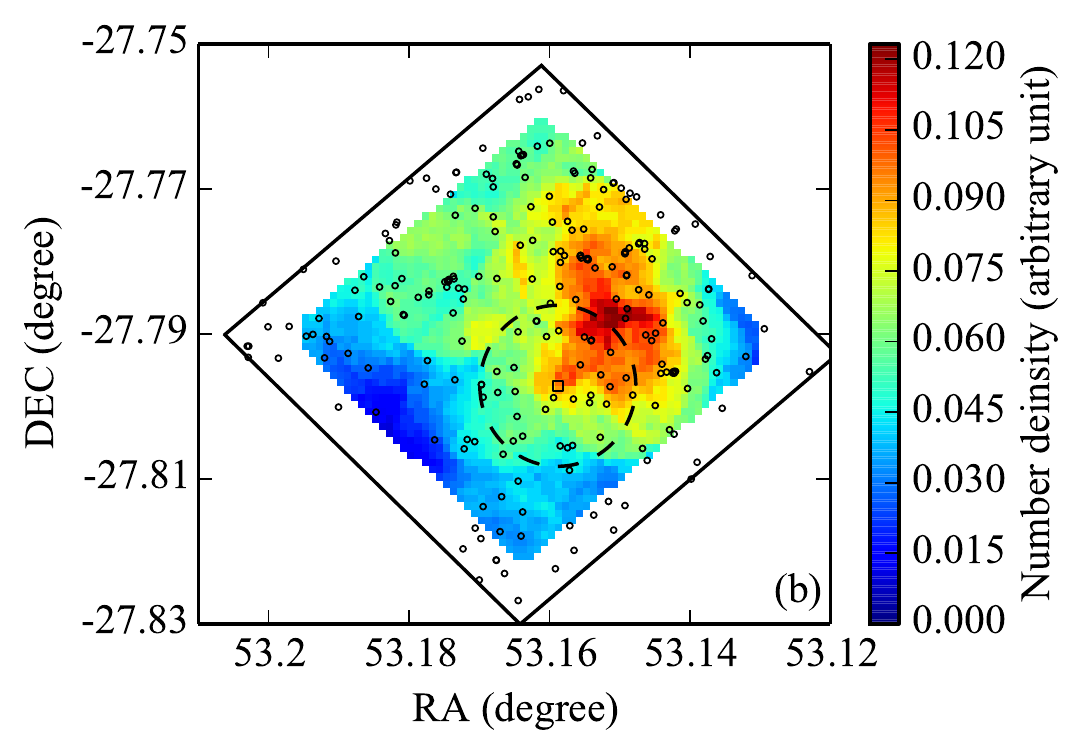}
\caption{
(a) Color image of the compact galaxy, XDF463 (in a red square), and 34 satellite galaxies (in green circles) within the virial radius, $r_{\rm vir}\sim300$~kpc (large dashed circle).
The color image is a composite of $HST$/F606W (blue), F125W (green) and F160W (red) bands.
(b) Contour map of the number density of galaxies at the same redshift (black circles) as XDF463 ($z\sim1.923$; black square) in the XDF field (large rectangle).
The number density, in arbitrary unit, is calculated with the galaxies within 300~kpc at each point.
The virial radius of XDF463 is shown with white dashed circle, in which we find 34 satellites.
}
\label{fig1}
\end{center}
\end{figure*}

Although such a very deep spectroscopy is observationally expensive yet, extremely deep imaging data of high spatial resolution by the $Hubble\ Space\ Telescope$ ($HST$) is helpful.
In this letter, we investigate the accretion phase of the two-phase scenario with the extremely deep multi-band imaging data in eXtreme Deep Field (XDF).
The data in the XDF enables us not only to discover the new satellite galaxies, but also to study their stellar population.
Counting the in-situ star formation, we can estimate the net increase of the stellar mass of the host galaxy after satellite merging.

Throughout the paper, we assume $\Omega_m$ = 0.3, $\Omega_\mathrm{\Lambda}$ = 0.7 and $H_0$ = 70 kms$^{-1}$Mpc$^{-1}$ for cosmological parameters and AB magnitude system (Oke \& Gunn~\citeyear{oke83}; Fukugita~{et al.}~\citeyear{fukugita96}).


\begin{deluxetable}{ccccc}\label{tab1}
\tabletypesize{\footnotesize}
\tablewidth{0pt}
\tablecaption{One-component S\'ersic parameters of XDF463, derived in F160W band image.}
\tablehead{
\colhead{ID} &
\colhead{$m_{\rm F160W}$} &
\colhead{$r_{e}$ (kpc)} &
\colhead{$n$} &
\colhead{$b/a$}
}
\startdata
\phn XDF463 & $\phn 22.10 $& 0.55 & 3.13 & 0.66
\enddata
\end{deluxetable}

\section{DATA AND RESULTS}\label{sec2}
\subsection{XDF and UVUDF Imaging Data}
We make use of the multi-band imaging data from XDF project (Illingworth {et al.}~\citeyear{illingworth13}).
The XDF team compiled all the data taken with 9 filter bands of $HST$/ACS and WFC3-IR in the HUDF field for 10.8 and 4.7 arcmin$^2$ for ACS, respectively.
The limiting magnitude for each band reaches $29.1$-$30.3$~mag with 0.\!\arcsec35 aperture, much deeper compared with the CANDELS deep observation ($\sim28$th mag).
In addition, we utilize the deep UV imaging data from the Ultraviolet Hubble Ultra Deep Field (UVUDF; Teplitz {et al.}~\citeyear{teplitz13}), which is taken with WFC3-UVIS for F225W, F275W and F336W bands, where $5\sigma$ limiting magnitude is 27.8, 27.7 28.3~mag for 0.\!\arcsec2 aperture, respectively.
The v2.0 data of Epoch~3 is used, where the post-flash CCD readout mode is applied.

\subsection{Host Compact Galaxy: XDF463}
For the selection of compact galaxies, we first make use of the catalog in Szomoru~{et al.}~(\citeyear{szomoru12}), which studied compact galaxies in the GOODS-South field with CANDELS near-infrared data.
In the catalog, we search compact galaxies in the region where both ACS and WFC3-IR data of XDF are available.
To minimize the uncertainty, we select those with spectroscopic redshift ($z_{\rm spec.}$), finding only one compact galaxy, XDF463.
The location of XDF463 and its virial radius are shown in a pseudo color image in Fig.~\ref{fig1}.
Fortunately, XDF463 resides in the central part of the XDF, where we can search its satellite galaxies impartially.

The structural parameters of XDF463 is derived with GALFIT (Peng et al.~\citeyear{peng10}), assuming single S\'ersic profile, in the same manner as Morishita et al.~(\citeyear{morishita14}).
The derived parameters (Table~1) are consistent with those in Szomoru et al.~(\citeyear{szomoru12}).

The stellar mass and star formation rate (SFR) of XDF463 are estimated with FAST code (Kriek~{et al.}~\citeyear{kriek09}), where we set the stellar population model of GALAXEV (Bruzual \& Charlot~\citeyear{bruzual03}), solar metallicity and Chabrier (\citeyear{chabrier03}) IMF.
The Calzetti dust law (Calzetti {et al.}~\citeyear{calzetti00}) is adopted in the range of $0\leq A_\mathrm{V}\leq3.0$~mag by step of 0.1, and age is set to the range from 0.1~Gyr to the age of the universe at the galaxy redshift.
Two star formation histories, exponentially-declining SFR($t$)~$\propto \exp(-t/\tau)$ and delayed SFR($t$)~$\propto t\times\exp(-t/\tau)$, are assumed.
It is known that the SED-based star formation rate is significantly affected by the model, while stellar mass is not (Wuyts~{et al.}~\citeyear{wuyts12}).
As such, we use the average SFR of both histories and refer to the difference as the error for each galaxy (see Morishita~{et al.}~\citeyear{morishita15}).
The error, however, hardly changes our results, as we see in the following section.
We also obtain the rest-frame $UVJ$ colors with the best-fit SED.
The physical parameters of XDF463 are summarized in Table~2.

From the stellar mass, $M_*\sim3.9\times10^{10}M_\odot$, we estimate its virial radius at the redshift to be $r_{\rm vir}\sim 300$~kpc, based on the numerical simulation by Oser~{et al.}~(\citeyear{oser10}).
One massive galaxy in Oser~{et al.}~(\citeyear{oser10}), M0209 in their Table~1, has comparable stellar mass at $z\sim2$ with XDF463, and therefore we refer it to the simulated galaxy as a model sample in the following discussion.
To check the consistency, we compare the stellar mass growth of M209 with an abundance matching model derived in Behroozi et al.~(\citeyear{behroozi13_770}).
The number density of the galaxies with similar mass of XDF463 is estimated to be $\sim8.9\times10^{-4}$~Mpc$^{-3}$ at $z\sim2$ in Fig.~3 of Behroozi et al.~(\citeyear{behroozi13_770}).
The number density of the population at $z\sim0$ would be $\sim6.2^{+0.2}_{-0.3} \times10^{-4}$~Mpc$^{-3}$, which corresponds to the population with $M_*\sim1.8^{+0.2}_{-0.6} \times10^{11}M_\odot$.
Since this stellar mass evolution is in good agreement with the one of M209, we conclude that the model galaxy traces the typical galaxies in the same mass range.

\begin{figure}
\figurenum{2}
\begin{center}
\includegraphics[width=9cm,bb=0 0 300 150]{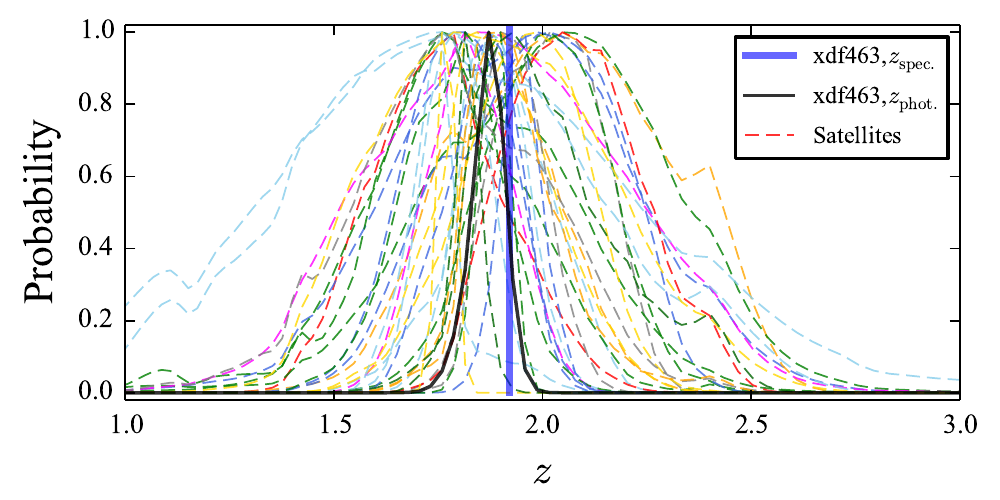}
\caption{
Probability distributions of photometric redshift for XDF463 (black solid line) and the satellites (colored dashed lines).
The spectroscopic redshift of XDF463 is shown with a blue vertical bar.
}
\label{fig2}
\end{center}
\end{figure}

\subsection{Satellite Galaxies around XDF463}
We investigate the satellite galaxies within $r_{\rm vir}$ of XDF463.
To identify the candidates of the satellite galaxies, we run SExtractor (Bertin\&Arnouts~\citeyear{bertin96}) in the ACS-all-combined image with a higher spatial resolution than WFC3-IR image.
The high spatial resolution of the ACS (pixel scale 30~mas, FWHM $\sim0.\!\arcsec11$), twice that of WFC3-IR (60~mas, $\sim0.\!\arcsec18$), is helpful to deblend small objects, less than $7\%$ of which are blended in WFC3-IR images.
The total number of the sources found within $r_{\rm vir}$ is 1369, while 324 of them are not detected in WFC3-IR image.
On the other hand, although 65 candidates in the WFC3-IR are not found in the ACS image, none of them is confirmed as the satellite of XDF463 with our criterion (see below).

By making use of the detection map by SExtractor, we conduct photometry on the 12-band images with pixel scale of 60~mas.
Each image used here is convolved to match F160W-band by the XDF team, and for the details we refer the readers to Illingworth~{et al.}~(\citeyear{illingworth13}).
Since SExtractor is known to overestimate the sky level, due to undetected faint galaxy outskirts, we do not use the count estimated with SExtractor.
Rather, we convolve the detection map with gaussian, and then estimate the sky background with unmasked pixels.

With 12 multi-band photometry, we estimate the photometric redshift ($z_{\rm phot}$) with EAZY (Brammer~{et al.}~\citeyear{brammer08}).
We evaluate the accuracy of our $z_{\rm phot}$ with spectroscopic redshifts ($z_\mathrm{spec}$) listed in the 3D-HST catalog (Skelton et al.~\citeyear{skelton14}).
Since galaxies with spectroscopic redshifts are biased to bright objects, we examine the redshift accuracy as follows.
First, we derive the photometric redshift of each galaxy using 3D-HST imaging data shallower than XDF data.
We then estimate the accuracy of redshift, $\delta z_{\rm crit}=\langle(z_{\rm phot} - z_{\rm spec})/(1+z_{\rm spec})\rangle$, for the galaxies at $1.5<z<2.0$ with $9.5<$log$M_*<10$.
The range of the stellar mass is chosen in order to see the galaxies around the limiting magnitude of the 3D-HST catalog ($H_{\rm F160W}\sim24.5$ or log$M_*\sim9.5$ at $z\sim2$; van der Wel et al.~\citeyear{vanderwel14}).
We find $\delta z_{\rm crit}$ and the fraction of the catastrophic redshift ($\delta z>0.5$, as defined in Kajisawa et al.~\citeyear{kajisawa11}) to be 0.051 and $14\%$, respectively.
By scaling the limiting magnitude of the 3D-HST catalog to that of the XDF data (F160W$\sim29.8$), we reach to the limiting stellar mass, log$M_*=7.2$.
Since the redshift accuracy should be comparable at each limiting magnitude for the 3D-HST and XDF data, we set $\delta z_{\rm crit}=0.051$ as the photometric accuracy for our sample with log$M_*>7.2$.
It is noted that there would be some systematic errors in the photometric redshift caused by varieties of SEDs, i.e. those for massive and less massive galaxies.
However, we hardly see any significant difference in the best-fit SEDs for both populations in our sample.
As our photometric accuracy is high enough for SED fitting, the photo-$z$ of our sample with low mass is well determined at $z\sim2$ (e.g., Kajisawa et al.~\citeyear{kajisawa11}; see also Bezanson et al.~\citeyear{bezanson15}).

We select the sample which satisfies $\delta z = (z_{\rm host} - z_{\rm phot})/(1+z_{\rm host})<\delta z_{\rm crit}$ as the satellite of XDF463, where $z_{\rm host}=1.9213$ is the spectroscopic redshift of XDF463.
Since some galaxies have a large error, we further exclude the sample with $\sigma_z/(1+z_{\rm phot})>\delta z_{\rm crit}$, where $\sigma_z$ is $1\sigma$ photometric redshift error derived by EAZY.
Out of the candidates, we find 35 galaxies in the criteria above.
One galaxy has spectroscopic redshift of $z=1.7672$, and is excluded from our satellite sample.
Finally we confirm 34 satellites within $r_{\rm vir}$ of XDF463 (see Fig.~\ref{fig1}a).
The redshift probability distribution of the satellites is shown in Fig.~\ref{fig2}, along with that of the host galaxy.
The physical properties of the satellites are obtained with FAST, in the same manner as done for XDF463, but fixed to the spectroscopic redshift of the host galaxy.
The results are summarized in Table~1.

We note that it is still uncertain whether the satellite galaxies defined with the above criteria are gravitationally bound to the host galaxy, because of the large uncertainty in photometric redshift ($\sim33000$~km s$^{-1}$).
This would be the general limitation of the photometric redshift, even with more accurate ones derived with medium-band imaging data (e.g., $\sim6000$-$10000$~km s$^{-1}$; Kawinwanichakij~{et al.}~\citeyear{kawinwanichakij14}).

The surface number density is calculated with the galaxies which satisfy the criteria above in all the XDF field.
At each point, we count the galaxies within 300~kpc, except for the peripheral region.
We verified the number density within $r_{\rm vir}$ of XDF463 is by a factor $\sim2.3$ higher than the median density in the XDF for the galaxies at the same redshift range.
It is worth noting that there is no massive galaxy at the redshift around XDF463, and hence it would be reasonable to assume that the satellite galaxies found in this study are bound to the galaxy.
The east neighboring region outside $r_{\rm vir}$ of XDF463 also has the peak at $z\sim1.9$, with the similar redshift distributions.
These galaxies could be also gravitationally bound to the same halo of XDF463 (see discussion below).


\begin{figure}
\figurenum{3}
\begin{center}
\includegraphics[width=8.5cm,bb=0 0 288 180]{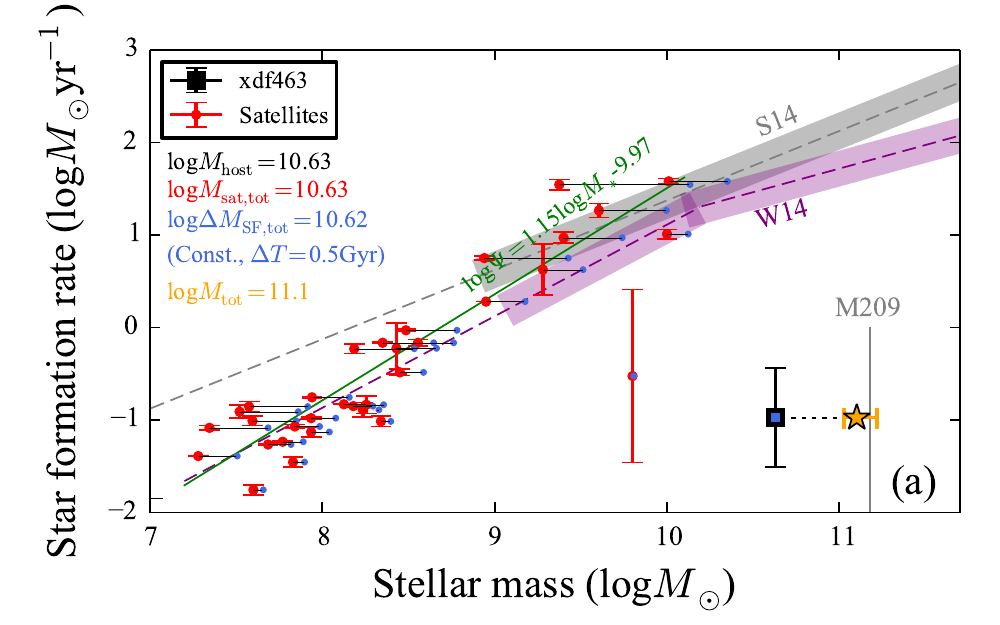}
\includegraphics[width=8.5cm,bb=0 0 288 180]{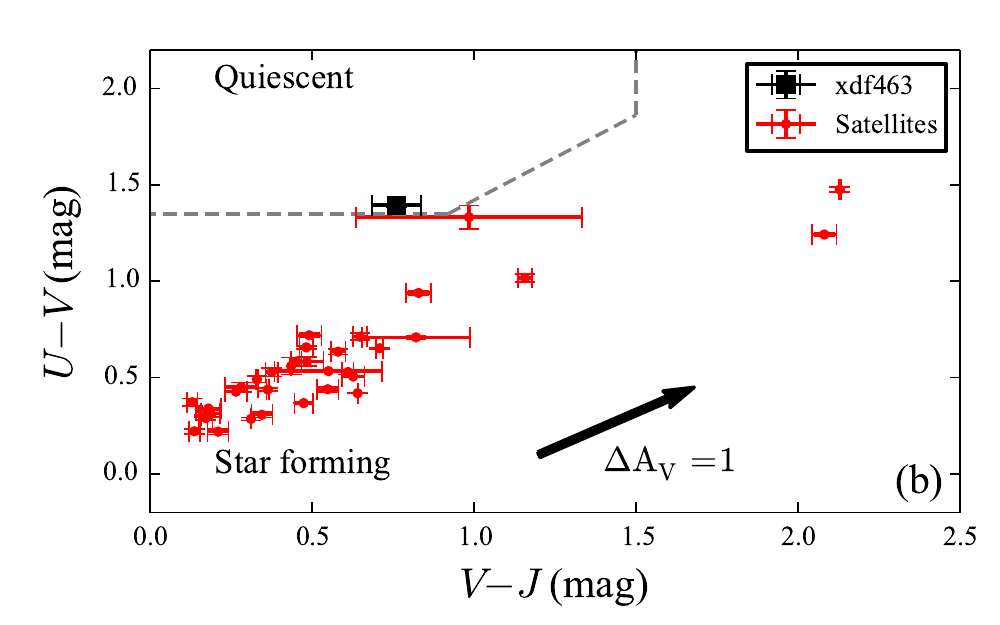}
\caption{
(a) Relation between the stellar mass and star formation rate (SFR) for XDF463 (black squares) and the satellites (red points), along with the main sequence of the star forming galaxies at $z\sim2$ by Speagle~{et al.}~(\citeyear{speagle14}) (S14; gray dashed line) and at $1.5<z<2$ by Whitaker~{et al.}~(\citeyear{whitaker14}) (W14; magenta dashed line), and our best fit slope, log~$\Psi=0.99$~log~$M_*-8.44$ (green solid line).
The shade of the slope for the previous results represents the typical uncertainties ($\sim0.2$~dex).
The expected stellar mass for each galaxy, by assuming the constant SFR for $\Delta T=0.5$~Gyr, is shown with a blue smaller point.
The total mass, if all satellites form stars for $\Delta T=0.5$~Gyr and accrete to XDF463, comes to an orange asterisk symbol.
The horizontal error bar of the asterisk accounts for the uncertainty of $\Delta T$, where we here set to 0.3~Gyr and 1.0~Gyr (see also the main text and Fig.~\ref{fig4}).
(b) XDF463 (black squares) and the satellites (red points) in the rest-frame $UVJ$ color diagram, where the quenched galaxies reside upper left region enclosed with dashed lines and star forming in the other region.
The error in the colors for each galaxy is estimated by the two different star formation histories.
The uncertainty in dust attenuation of $\Delta A_{\rm V}=1.0$~mag is shown with an arrow.
}
\label{fig3}
\end{center}
\end{figure}

\section{Discussion}\label{sec3}

\subsection{Total-Mass Increase by Satellite Accretion}\label{sec31}
We first investigate the stellar mass and SFR of the satellites in Fig.~\ref{fig3}a.
We find the satellites having stellar mass $\sim10^{10} M_\odot$ down to $1.6\times10^7 M_\odot$.
The total mass of 34 satellites is $\sim 4.3\times10^{10}M_\odot$, which is comparable to that of XDF463.
The mass growth of XDF463 by the accretion of the satellites is, however, in deficit to become a massive elliptical galaxy in the local universe ($\sim10^{11}M_\odot$) expected from the numerical simulations (e.g., Oser~{et al.}~\citeyear{oser10}) and the observations (e.g., van Dokkum~{et al.}~\citeyear{vandokkum10}), even if all satellites fall to the central galaxy.
The mass assembly history of M0209 in Oser~{et al.}~(\citeyear{oser10}) shows that $\sim20\%$ of the total stellar mass at $z=0$ ($M_0\sim1.5\times10^{11}M_\odot$) is formed at $z>2$ with in-situ mechanisms such as cold gas accretion and collapse-like star formation, while $\sim80\%$ is formed at or outside the virial radius and accretes to the central galaxy at $z<2$.
On the other hand, the total stellar mass of the satellites found in this study accounts only for $\sim30\%$ of $M_0$.

In Figs.~\ref{fig3}a and b, most satellites found in this study are classified as star forming and form stars at the comparable rate to the main sequence of the star forming galaxies at the same redshift.
The relations for the main sequence galaxies at $z=2$ of Speagle~{et al.}~(\citeyear{speagle14}; S14) and at $1.5<z<2.0$ of Whitaker et al.~(\citeyear{whitaker14}; W14) are shown as reference, while our estimate for the slope is $1.15$ (see the following section for the details).
The in-situ star formation of the satellites should  contribute to the mass increase of XDF463 during their infall to the central galaxy.
Here we estimate the stellar mass due to the in-situ star formation for each satellite by assuming the same SFR over $\Delta T=0.5$~Gyr and mass return fraction of $R=0.36$, i.e. $\Delta M_{\rm SF} = SFR \times (1-R)  \times\Delta T$.
The expected increase of stellar mass for each satellite is shown with blue points in Fig.~\ref{fig3}a.
The total stellar mass made by the in-situ star formation of the satellites would be $\sim4.3\times10^{10}M_\odot$, comparable to the $already\ existing$ stellar mass of the satellites.
Counting the mass growth by the in-situ star formation, we find the total mass of the accreting satellites and XDF463 could be $\sim1.3\times10^{11}M_\odot$, $\sim83\%$ of $M_0$, as shown in Fig.~\ref{fig3}a.
Although our estimation still lacks $\sim17\%$ of the mass to the expectation, we note that there are uncertainties in $\Delta T$ and the total number of the satellites, i.e. the uncertainty in the photometric redshift.
The duration of the star formation is one of the most unknown parameters because of complex physics in massive halos in merging events.
As such, we also show the mass growth estimated with $\Delta T=0.3$~Gyr (which results in $71\%$ of $M_0$) and $1.0$~Gyr ($110\%$; as shown with errorbars in Fig.~\ref{fig3}a).
The merger time scale estimated with the numerical calculation is typically $\sim1.0$-1.5~Gyr for the satellite at $\sim100$~kpc (Lotz~{et al.}~\citeyear{lotz11}), or longer (see below), and we here set $\Delta T=1.0$~Gyr as the maximum time duration of the star formation.
In Fig.~\ref{fig3}a we show the mass growth of XDF463 with constant SFR, while during the merger event it significantly varies, though is unpredictable with the present data.
We also note that there is uncertainty in the stellar mass (a factor of $\sim1.4$) when considering the similar halo mass in the same simulation (see Oser et al.~\citeyear{oser10}).

We show the rest-frame colors in the $UJV$ color diagram in Fig.~\ref{fig3}b, to investigate the star formation activity (Williams~{et al.}~\citeyear{williams09}; Morishita~{et al.}~\citeyear{morishita14}).
Most of the satellites are classified to be star forming galaxies, as expected in Fig.~\ref{fig3}a, and two at the rightmost are significantly dust attenuated, while the host compact galaxy is in the quenched region.
The blue rest-frame colors of the satellites suggest that they are not quenched yet even in the massive halo (see discussion below).

It is noted that the accretion of satellites with gas (wet merger), on the contrary, would suppress the size growth (Naab~{et al.}~\citeyear{naab09}), while some studies suggest the necessity of wetness ($\sim4\%$ of gas to the total stellar mass; Sonnenfeld~{et al.}~\citeyear{sonnenfeld14}) in the context of the halo mass profile.
Nevertheless, we here stress that the in-situ star formation of satellites would increase the stellar mass in the accreting phase, which would explain the significant mass growth at the outer envelope and supplement the deficit of the accreting component.

\begin{figure*}
\figurenum{4}
\begin{center}
\includegraphics[width=8cm,bb=0 0 288 288]{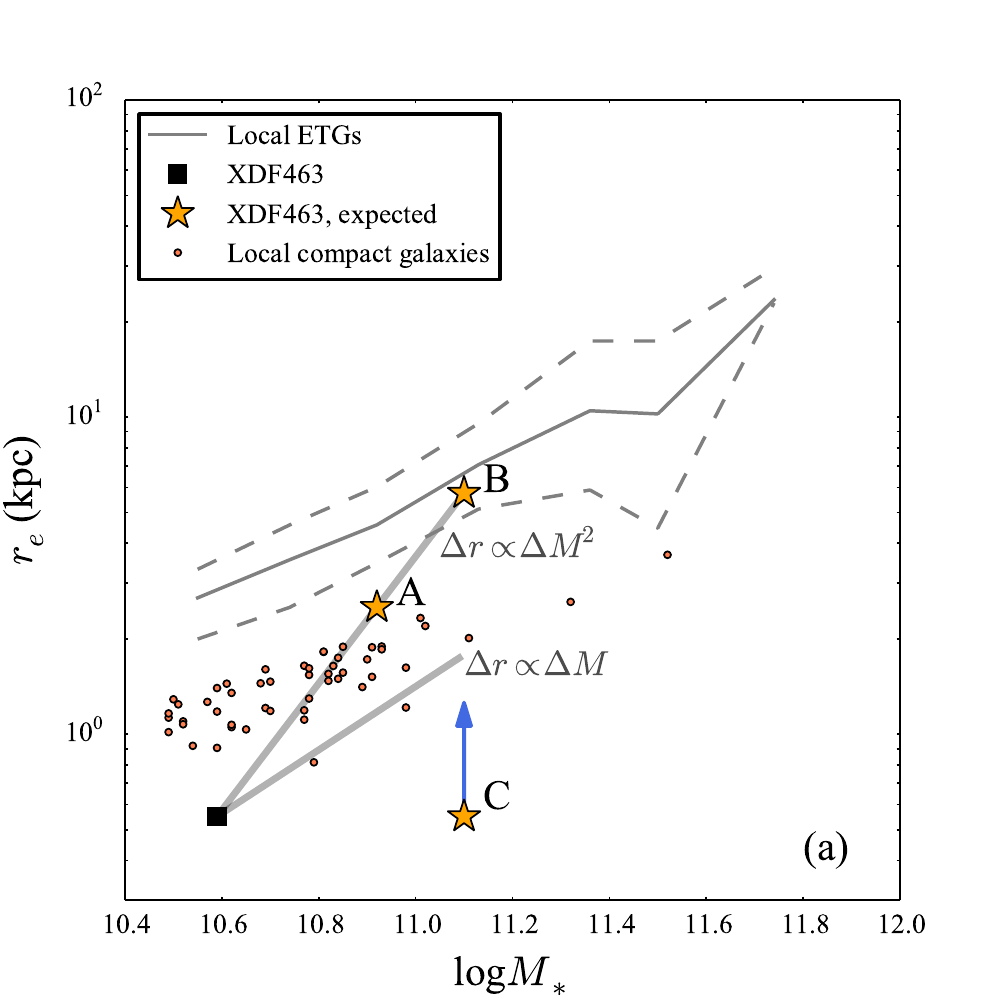}
\includegraphics[width=8cm,bb=0 0 288 288]{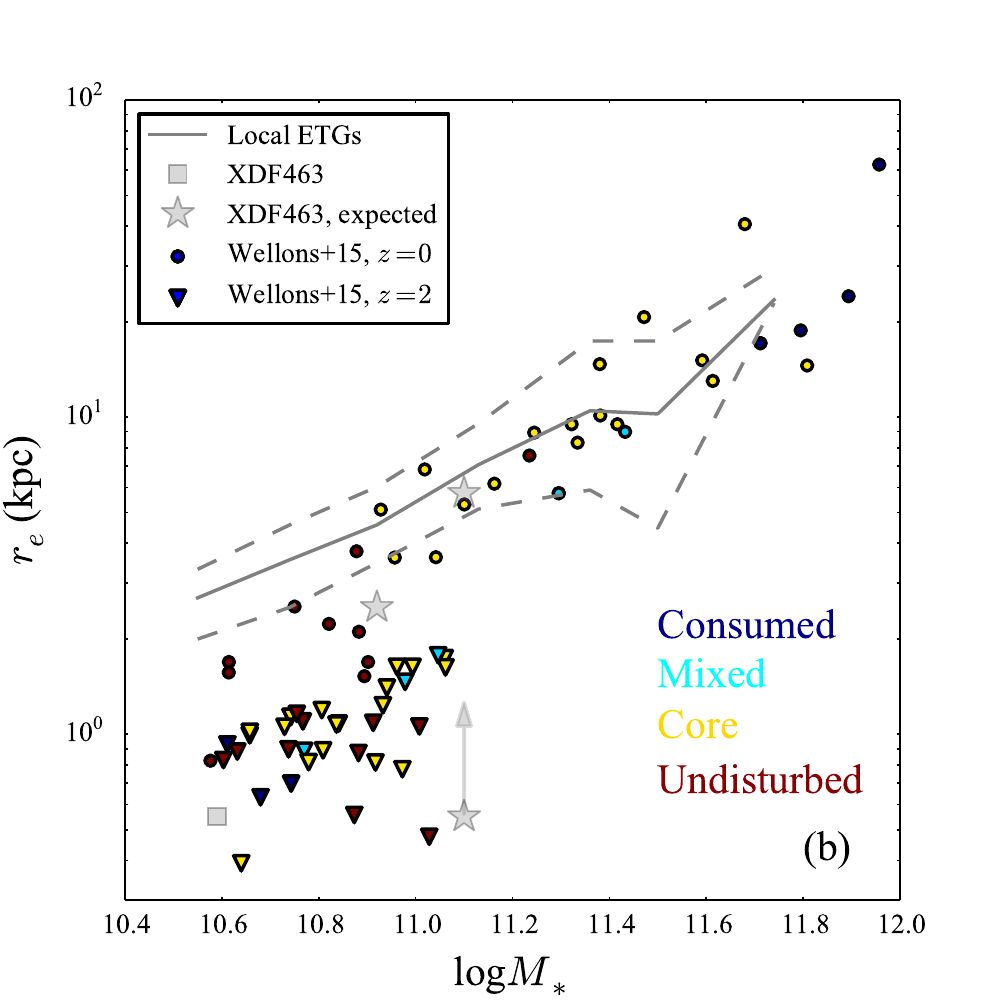}
\caption{
(a) Observed and expected size-stellar mass relation for XDF463 (black squares and yellow stars, respectively).
The relation for the local early-type galaxies (Poggianti et al.~\citeyear{poggianti13}) are shown (gray solid lines) with 1~$\sigma$ error (dashed lines).
Three predictions are shown, based on different assumptions for the satellite fate; they quench very soon and accrete to the outer envelope of the host (A), continue in-situ star formation and then quench before they accrete (B), or continue forming stars until they induce wet mergers (C, lower limit).
Compact galaxies in the local universe (Poggianti et al.~\citeyear{poggianti13}) are shown with red circles.
The evolution via major ($\Delta r\propto \Delta M$) and minor ($\Delta r\propto \Delta M^2$) mergers expected by assuming the virial theorem (e.g., Hopkins et al.~\citeyear{hopkins09b}) are shown with arrows.
(b) Compact galaxies at $z=2$ (inverted triangles) and their progenitors at $z=0$ (circles) in the Illustris simulation (Wellons et al.~\citeyear{wellons15}) are shown.
Each data point is color-coded by the evolution path (see the main text).
It is noted that the Illustris data are arbitrarily shifted for the comparison with our observed and expected values for XDF463 (shown with gray square and stars).
}
\label{fig4}
\end{center}
\end{figure*}

\subsection{Stellar Mass-Size Relation}\label{sec32}
We show the stellar mass-size diagram of XDF463 and its satellites in Fig.~\ref{fig4}a, along with that of the local early-type galaxies derived in Poggianti et al.~(\citeyear{poggianti13}).
We convert the stellar mass in Poggianti et al.~(\citeyear{poggianti13}), which is derived with Kroupa (\citeyear{kroupa01}) IMF, to one with Chabrier IMF, by using the relation in Cimatti et al.~(\citeyear{cimatti08}).
Considering the mass growth expected in the previous section, we find that XDF463 should prefer minor merger, if it significantly grows its size to be on the local relation ($r_{e}\sim5$~kpc), with the relation of $\Delta r\propto \Delta M^2$ derived with virial theorem (e.g., Naab et al.~\citeyear{naab09}), rather than major merger with $\Delta r\propto \Delta M$.
However, the significant size growth expects merging with quenched satellites (dry minor merger), contrary to the finding in this study.
Because of the complex dissipative process, it would not be plausible if wet merger significantly increases the galaxy size as well as minor dry merger.
In the following, we pursue three extreme, but possible, scenarios for the mass and size growth of XDF463, focusing on the satellite properties.
The expected mass and size growth of XDF463 should be within the scope of these three cases.

We first investigate the size growth via purely dry minor merger (Case~A), neglecting the in-situ star formation of the satellite galaxies, as is done in previous studies: the gas in the satellites is quickly exhausted or tidally stripped (e.g., environmental quenching; see also Section 3.4), and then the satellites accrete into the outer envelope of the host galaxy in due time.
In this scenario the size evolution follows $\Delta r\propto \Delta M^2$ (A in Fig.~\ref{fig4}a).
Considering the mass and size evolution in Fig.~\ref{fig4}a, we expect that the compact galaxy ends in the population of local compact galaxies (Poggianti et al.~\citeyear{poggianti13}), without any wet compaction process (see below).
It should be reminded that we find a dense region at the north-east of XDF463 (Fig.~\ref{fig1}b).
It is possible that some galaxies outside the virial radius of XDF463 would fall and then increase the mass and size, which would bring the compact galaxy on the local size-mass relation. 

When counting the in-situ star formation of the satellites, the situation is dramatically changed.
As shown in the previous section, we now take into account the mass growth of the satellites and assume that they stop star formation activity before merging to the host (Case~B).
In this scenario, we expect again the size growth with $\Delta r\propto \Delta M^2$.
Since stellar mass growth is much larger, the size growth is accordingly more significant than one in Case~A, to be consistently on the local relation in Fig.~\ref{fig4}a.

It is possible that satellites continue star formation activity until they merge to the host (Case~C).
However, it would be more difficult to predict the size evolution because of the complicated physical mechanisms.
The host galaxy significantly changes its morphology to, e.g., the late-type galaxy with much larger scale radius, or, on the opposite case, evolves to compact galaxy through the wet compaction (e.g., Dekel\&Burkert~\citeyear{dekel14}).
As such, we show the expected mass and size in Fig.~\ref{fig4}a.

It is also worth mentioning that if XDF463 experiences merger with the largest gas-rich satellites, it could end up with one of the local compact galaxies, by following $\Delta r\propto \Delta M$.
Since major merger typically involves dissipative process and hence lose the angular momentum, the compact stellar system is a possible end-product (e.g., Wuyts et al.~\citeyear{wuyts10}).

More specific discussion is available when we take account of results by numerical simulations.
In Fig.~\ref{fig4}b, we show the compact galaxies at $z=2$ and their descendants at $z=0$, taken from the Illustris simulation (Wellons et al.~\citeyear{wellons15}).
Since the compact galaxies in their study are more massive (log$M_*>11$), we here focus to discuss the relative evolution, rather than the specific position on the mass-size relation.
We multiply their 3-dimensional effective radius by 0.75, to fairly compare with the observed 2-dimensional effective radius.
In Wellons et al.~(\citeyear{wellons15}), they classified the fate of the compact galaxies as follows: Consumed (absorbed by other more massive galaxies), Mixed (major merger), Core (minor merger, which keeps the compact galaxy), and Undisturbed.
As mentioned above, minor merger evolves the galaxy size to the local relation more efficiently compared to major merger.
Some of the satellites involved in minor merger might have gas, although we could not identify them only with the data in Wellons et al.~(\citeyear{wellons15}).

\subsection{Star Formation Main Sequence of Satellite Galaxies}
We here investigate the relation between stellar mass and SFR (star formation main sequence) of the satellite galaxies found in this study, as done for massive galaxies in the local universe (e.g., Elbaz et al.~\citeyear{elbaz07}, \citeyear{elbaz11}; Zahid et al.~\citeyear{zahid12}) and at high-$z$ (e.g., Daddi et al.~\citeyear{daddi07}; Kajisawa et al.~\citeyear{kajisawa10}; Suzuki et al.~\citeyear{suzuki15}).
First of all, we note that the SFR derived through SED fitting depends on the assumed star formation history, while one derived from nebular lines, e.g., H$\alpha$, is suffered from dust extinction.
Because the systematic difference is much larger than other uncertainties, e.g., errors in photometry, we thus estimate the uncertainty in the SED-based SFR by assuming two star formation histories described in Section~2.2.

In Fig.~\ref{fig3}a, we show the best fit slope for the satellite galaxies in the form of,
\begin{equation}
\mathrm{log} \Psi = a\ \mathrm{log} M_* + b,
\end{equation}
where $\Psi$ denotes SFR, and best fit values of $a=1.15\pm0.02$ and $b=-9.97\pm0.20$.
We exclude from the fit one satellite galaxy near $UVJ$ boundary because it could be classified as quiescent within its error (Fig.~\ref{fig3}b).
We also show two slopes by Speagle et al.~(\citeyear{speagle14}) at $z\sim2$ (S14) and Whitaker et al.~(\citeyear{whitaker14}) at $1.5<z<2.0$ (W14).
The best fit slope derived in the present study is much steeper than those in the previous studies ($a\sim0.6$).
Although W14 suggests a steeper slope ($a=1.04 \pm 0.05$) at lower mass range (log$M_*<10.2$), the inconsistency could not be explained within the uncertainties in SFR and stellar mass.
This is mainly because of the low-mass population with log$M_*\lesssim8$, which is not included in the previous studies at the redshift range.
Interestingly, the main sequence slope becomes much milder ($a=0.71\pm0.09$) if derived only for more massive galaxies (log$M_*>8.8$), which is in good agreement with S14 ($a=0.75\pm0.03$), suggesting a possible knee at lower mass than of that of W14 (but see also Abramson et al.~\citeyear{abramson14}).
It is noted that since the satellite galaxies discussed here is above stellar mass completeness (both for star forming and quiescent populations), the best fit value for the slope is not biased.
Further detection of any galaxies with lower star formation rate under the detection limit would change the derived slope much steeper.
Uncertainties in redshift and SED parameters (star formation rate and stellar mass) hardly change our result, because they should affect all the mass range and, therefore, only give the offset.
Therefore, we conclude that the result of the steeper slope at the low mass range is robust.
It is worth noting that the main sequence slope for $\Delta M$-evolved galaxies (blue points in Fig.~\ref{fig3}a) would become much milder than the current slope and become naturally consistent with the previous results, which might support our assumption of the constant star formation during $\Delta T$.

\subsection{Quenching of Satellite Galaxies}\label{sec34}
We have estimated the mass growth due to the in-situ star formation of the satellites with a simple assumption: star formation of each satellite is kept at constant rate for $\Delta T$.
Although the assumption is reasonable for galaxies in the field environment (Genzel~{et al.}~\citeyear{genzel10}; Kennicutt \& Evans~\citeyear{kennicutt12}), it is not clear for the satellites in the massive halo of XDF463, where some environmental effects are expected (Peng~{et al.}~\citeyear{peng10}; see also Koyama~{et al.}~\citeyear{koyama14} for the high-$z$ cluster environments).
We find no significant dependence of the SFR nor $UVJ$ colors of the satellites on the distance from the center, i.e. no evidence of environmental quenching is observed.
Although the fact that the star formation activity of the satellites is independent of the distance from the central galaxy is consistent with the result of Wetzel~{et al.}~(\citeyear{wetzel13}), who have found the satellite galaxy continues its star formation for several Gyr after the infall within $r_{\rm vir}$ of the host galaxy, it is noted that most of the satellites found in this study reside at $r>100$~kpc.
As shown in Oser~{et al.}~(\citeyear{oser10}), most satellites in the massive halo collapse at far from the center, $r>r_{\rm vir}$, at the higher redshift.
Considering that we have few satellites close to the host galaxy, it is likely that we are witnessing the satellites right after their infall wihtin $r_{\rm vir}$.
The result of the blue rest-frame colors of the satellites in Fig.~\ref{fig3}b would also support this idea.
We note that some galaxies might be quenched not only by the environmental but also by other mechanism related to, e.g., galaxy structure (morphological quenching; Martig et al.~\citeyear{martig09}; Genzel~{et al.}~\citeyear{genzel14}; Kajisawa~{et al.}~\citeyear{kajisawa15}).

It should be recalled that the local elliptical galaxies have rather flat age gradient, $<1$~Gyr difference between inner and outer radii (e.g., Tamura et al.~\citeyear{tamura00}), suggesting that these satellites would be quenched soon if the compact galaxy is the progenitor of the local elliptical galaxy.
However, as mentioned in Section~3.2, there is alternative evolution path to massive late-type galaxies and compact galaxies, when it experiences, e.g., gas-rich merger (Case C).

\begin{figure*}
\figurenum{5}
\begin{center}
\includegraphics[width=12cm,bb=0 0 288 180]{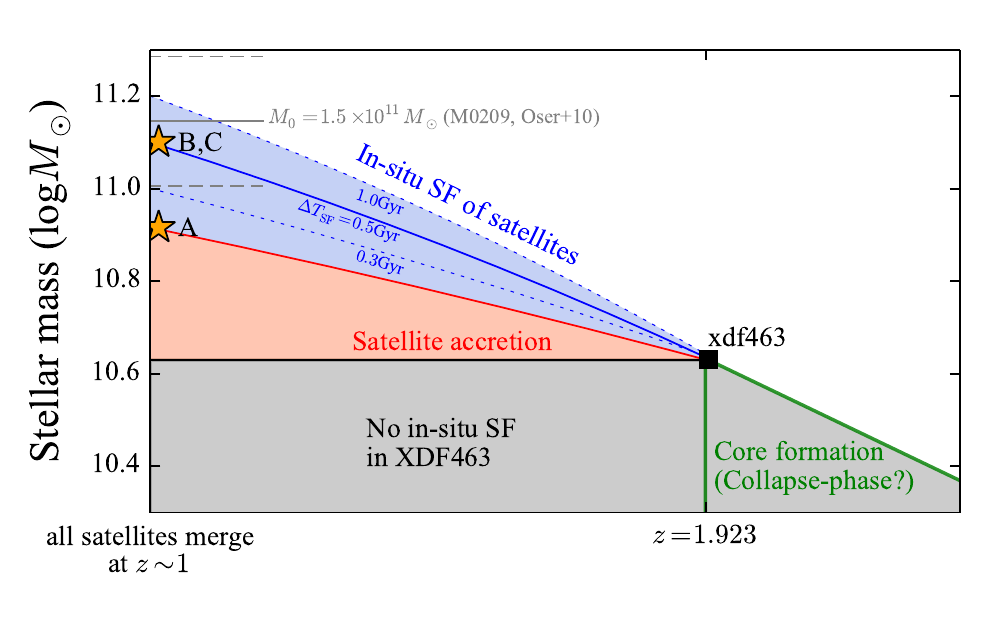}
\caption{
Summary figure of the expected growth of stellar mass as a function of redshift.
The compact galaxy, XDF463, has $\sim23\%$ of the stellar mass expected at $z=0$, $M_0=1.5\times10^{11}M_\odot$ (M0209 in Oser~{et al.}~\citeyear{oser10}; horizontal gray solid line) at the observed redshift.
The systematic uncertainty in $M_0$ is shown with horizontal dashed lines (see the main text).
The satellites found in this study account for $\sim26\%$ of $M_0$.
The in-situ star formation of the satellites are assumed with $\Delta T=0.5$~Gyr (blue solid line), which accounts for $\sim28\%$ of $M_0$, and net mass increase by the satellites would be $\sim54\%$ of $M_0$ if all of them are merged in typical dynamical time ($\sim1$~Gyr).
The contribution of the in-situ star formation of the satellites with different time duration ($\Delta T=0.3$~Gyr and 1.0~Gyr) is also shown with dotted blue lines.
}
\label{fig5}
\end{center}
\end{figure*}

\subsection{Fate of XDF463}
So far we have investigated the mass increase of the compact galaxy, XDF463, with the accreting satellites taking account for their in-situ star formation.
According to the previous observational and theoretical studies, these satellites would merge to the central galaxy in several Gyr, except for the smallest satellites ($<1/100$ of the host galaxy), whose accretion time would exceed the Hubble time (Lotz~{et al.}~\citeyear{lotz11}; Wetzel~{et al.}~\citeyear{wetzel13}; Tal~{et al.}~\citeyear{tal14}).
In other words, the compact galaxy observed at $z\sim2$ would be settled as the local analogous at $z\sim1$, as shown in many studies of massive galaxies (e.g., Bundy~{et al.}~\citeyear{bundy09}).
This is also consistent with the recent study on the evolution of the number density of compact galaxies (van Dokkum~{et al.}~\citeyear{vandokkum15b}; but see also Damjanov~{et al.}~\citeyear{damjanov15}).
We show the summary view of the present study in Fig.~\ref{fig4}.

After all satellites we found in this study merge to the host galaxy, does not any merger happen?
This seems not to be the case; some previous studies show, by counting the pair fraction, that the merger rate of massive galaxies is still nonnegligible at $z<1$ ($\sim20\%$; Bluck~{et al.}~\citeyear{bluck12}; Newman~{et al.}~\citeyear{newman12}).
Although we have investigated satellites within $r_{\rm vir}$ of XDF463, it is noted again that some galaxies would fall into $r_{\rm vir}$ at the later epoch (Tal~{et al.}~\citeyear{tal14}).
Furthermore, satellite galaxies typically have eccentric orbit (van den Bosch~\citeyear{vandenbosch99}; see also Hayashi \& Chiba~\citeyear{hayashi14}), and avoid being detected within the projected $r_{\rm vir}$ at the observed redshift.
As noted before, we have found a peak of the number density of the galaxies at $z\sim1.9$ in the neighboring region at $r>r_{\rm vir}$ (Fig.~\ref{fig1}b), whose redshift distributions are similar to those of XDF463.
Since there is no nearby massive galaxy comparable to XDF463, the galaxies might be trapped within the host halo of XDF463, though further spectroscopic observation would be required.

\subsection{Caveats}
We note caveats on our new finding that most satellite galaxies are star forming and reside at $r>100$~kpc from the compact galaxy.
The finding would be inconsistent with the results of previous studies in some aspects.
For example, Newman et al.~(\citeyear{newman12}) searched satellite galaxies in $10$~kpc~$<r<30$~kpc around massive quiescent galaxies at $z\sim2$, and found that $\sim38\%$ of the satellites are quiescent, though their sample is a mixture of compact and normal quiescent galaxies.
We recall two problems to be elucidated: the completeness of the low mass quiescent galaxies and the photometric redshift uncertainty.

Firstly, we investigate the detection limit for low mass quiescent satellites.
Since quiescent galaxies are more difficult to detect than star forming galaxies with the same stellar mass, the detection bias should be properly examined.
As described in Section~2, we set the stellar mass limit by scaling that of van der Wel et al.~(\citeyear{vanderwel14}) to the detection limit of XDF data (log$M_*>7.2$).
The stellar mass limit in van der Wel et al.~(\citeyear{vanderwel14}) is for quiescent galaxies.
Since we have no quiescent satellite even in more massive stellar mass range, shifting the mass limit does not change the fraction of quiescent satellite.
Furthermore, we investigate the satellite galaxies around massive quiescent galaxies (log$M_*>10.6$), but not compact, at $1<z<2.5$ in the XDF, in the same manner as for XDF463.
We find that $\sim50\%$ of the satellites are quiescent if the same mass limit as Newman et al.~(\citeyear{newman12}), log$M_*>9.5$, is applied.

We then investigate the possibility that the quiescent galaxies are overlooked due to the redshift error.
As an extreme test, we fix the redshift of all the detected candidates within $r<300$~kpc to the same redshift of the host ($z=1.92$), and derive SED parameters and the rest-frame colors of candidates.
Intriguingly, we find no quiescent galaxies among them, strongly supporting our finding that most satellites around the compact galaxy are blue star forming galaxies.

We therefore conclude that XDF463 has some extraordinary nature when compared to the non-compact massive quiescent galaxies at the comparable redshift, as mentioned in Sections~3.1 and 3.4; surrounded by numerous star forming satellites, whose distances are mostly $>100$~kpc.
The finding would give us an important clue to the formation and fate of compact galaxies, which we further investigate in the future work.


\section{CONCLUSION}\label{sec4}
In this paper, we investigate the latter epoch of the ``two-phase" scenario, the accretion phase, by searching the accreting satellites within the virial radius of the compact galaxy, XDF463. 
Several previous studies have observationally challenged the scenario in the similar way, resulting in difficulty to explain the total mass increase only with the $existing$ satellites found in the near-infrared imaging.
For the further investigation, we made use of the extremely deep imaging data ($m_{\rm ACS}\sim30.6$~ABmag) taken with $HST$/ACS, which enabled us to find more faint satellites.
Some of them are not detected with $HST$/WFC3-IR, in spite of the comparable imaging depth, partly because of the confusion limit, and so does the previous ground-based near-infrared imaging survey.
For the photometrically confirmed 34~satellite galaxies out of 1369 candidates within $r_{\rm vir}$, we derived their SED properties i.e. SFR and stellar mass, finding that most of them are classified as star forming in the $UVJ$ diagram.
However, low-mass satellites are significantly below the star formation main sequence, with steeper slope of log~$\Psi=1.15$~log~$M_*-9.97$ than those in the previous studies (Fig.~\ref{fig3}a).

By assuming all of the satellites would merge in $\sim0.5$~Gyr, while keeping in-situ star formation (B in Fig.~\ref{fig4}a) the estimated mass increase due to the accretion and in-situ star formation would be $\sim8.3\times10^{10}M_\odot$, which is roughly consistent with the expectation of the observations and simulations within uncertainties.
We, however,  do not give a definitive conclusion how the size of the compact galaxy evolves, because of insufficient information for the satellite properties, which leads to the other evolution paths (A and C in Fig.~\ref{fig4}a).
Since there are many compact galaxies found in the local universe, we stress that the significant size evolution to a local elliptical galaxy might not be only the fate of XDF463.
Although the uncertainties in the duration of the star formation, merging time scale of the satellites, and photometric redshift affect the estimate, the contribution of the in-situ star formation of the $satellites$, which has been ignored in previous studies, would mitigate the known difficulty on the significant size and mass evolution of compact galaxies to local massive galaxies.

\acknowledgments
We thank an anonymous referee for the valuable and constructive comments, and Sarah Wellons for kindly providing the structural parameters of compact galaxies in the Illustris simulation.
This work has been financially supported by a Grant-in-Aid for Scientific Research (24252203) of the Ministry of Education, Culture, Sports, Science and Technology in Japan. 
T.M. acknowledges support from the Japan Society for the Promotion of Science (JSPS) through JSPS research fellowships for Young Scientists.

{\it Facilities:} \facility{$HST$ (ACS, WFC3)}.

\begin{deluxetable*}{clccccccccc}\label{tab2}
\tabletypesize{\footnotesize}
\tablewidth{0pt}
\tablecaption{Physical properties of XDF463 and the Satellite Galaxies}
\tablehead{
\colhead{ID} &
\colhead{$z_{\rm phot}$\tablenotemark{b}} &
\colhead{RA} &
\colhead{DEC} &
\colhead{$r_{\rm proj}$\tablenotemark{c}} &
\colhead{log$M_{\rm *}$\tablenotemark{d}} &
\colhead{logSFR\tablenotemark{d}} &
\colhead{$U-V$\tablenotemark{d}} &
\colhead{$V-J$\tablenotemark{d}}
}
\startdata
463\tablenotemark{a} & $\phn 1.87 _{-0.05}^{+0.10}$& 53.158807 & -27.797155 & -- & 10.59 & -1.51 & 1.39 & 0.69 \\
2765 & $\phn 1.87 _{-0.06}^{+0.19}$& 53.157426 & -27.805705 & 268 & 8.06 & -0.83 & 0.43 & 0.23 \\
2819 & $\phn 2.02 _{-0.11}^{+0.22}$& 53.156708 & -27.805386 & 262 & 7.89 & -0.76 & 0.30 & 0.16 \\
2823 & $\phn 1.79 _{-0.17}^{+0.25}$& 53.158469 & -27.805454 & 258 & 7.57 & -0.96 & 0.28 & 0.19 \\
2995 & $\phn 1.99 _{-0.29}^{+0.46}$& 53.165155 & -27.804768 & 294 & 7.84 & -1.40 & 0.65 & 0.50 \\
3073 & $\phn 2.05 _{-0.24}^{+0.47}$& 53.152766 & -27.804261 & 276 & 7.48 & -0.98 & 0.21 & 0.18 \\
3074 & $\phn 1.93 _{-0.04}^{+0.21}$& 53.163815 & -27.804110 & 256 & 8.44 & -0.04 & 0.21 & 0.12 \\
3756 & $\phn 2.05 _{-0.24}^{+0.43}$& 53.164614 & -27.801389 & 207 & 7.62 & -1.26 & 0.56 & 0.43 \\
4026 & $\phn 1.76 _{-0.13}^{+0.41}$& 53.160554 & -27.800407 & 112 & 7.57 & -1.81 & 0.73 & 0.45 \\
4235 & $\phn 1.76 _{-0.13}^{+0.25}$& 53.151869 & -27.799679 & 206 & 7.54 & -0.91 & 0.29 & 0.31 \\
4253 & $\phn 1.87 _{-0.07}^{+0.17}$& 53.154280 & -27.799488 & 144 & 8.93 & 0.28 & 0.65 & 0.70 \\
4459 & $\phn 1.84 _{-0.16}^{+0.42}$& 53.156591 & -27.798986 & 83 & 7.93 & -1.18 & 0.73 & 0.63 \\
4466 & $\phn 2.02 _{-0.13}^{+0.24}$& 53.159414 & -27.798794 & 53 & 7.90 & -0.99 & 0.43 & 0.23 \\
4471 & $\phn 1.81 _{-0.07}^{+0.17}$& 53.154104 & -27.798459 & 135 & 8.52 & -0.13 & 0.43 & 0.36 \\
4504 & $\phn 1.90 _{-0.07}^{+0.22}$& 53.169437 & -27.798734 & 296 & 8.31 & -0.93 & 0.60 & 0.44 \\
4605 & $\phn 1.81 _{-0.10}^{+0.18}$& 53.148138 & -27.798438 & 296 & 8.40 & -0.50 & 0.70 & 0.65 \\
4706 & $\phn 2.02 _{-0.13}^{+0.24}$& 53.167358 & -27.798077 & 237 & 8.31 & -1.07 & 0.95 & 0.79 \\
4769 & $\phn 1.84 _{-0.07}^{+0.25}$& 53.164888 & -27.797944 & 169 & 8.26 & -0.97 & 0.60 & 0.44 \\
4990 & $\phn 1.90 _{-0.05}^{+0.16}$& 53.169679 & -27.796962 & 299 & 10.00 & 0.96 & 1.04 & 1.14 \\
5007 & $\phn 1.79 _{-0.28}^{+0.43}$& 53.151370 & -27.797274 & 204 & 7.34 & -1.11 & 0.29 & 0.31 \\
5088 & $\phn 1.81 _{-0.12}^{+0.19}$& 53.169686 & -27.797023 & 299 & 9.39 & 1.49 & 1.25 & 2.04 \\
5339 & $\phn 2.05 _{-0.11}^{+0.25}$& 53.149079 & -27.796065 & 269 & 8.18 & -0.86 & 0.51 & 0.33 \\
5408 & $\phn 1.96 _{-0.07}^{+0.12}$& 53.152687 & -27.795644 & 174 & 9.25 & 0.35 & 0.53 & 0.38 \\
5515 & $\phn 1.90 _{-0.07}^{+0.11}$& 53.167488 & -27.795190 & 246 & 9.37 & 0.91 & 0.43 & 0.52 \\
5883 & $\phn 1.76 _{-0.27}^{+0.50}$& 53.165050 & -27.794641 & 188 & 6.84 & -1.90 & 0.34 & 0.14 \\
5927 & $\phn 1.99 _{-0.16}^{+0.26}$& 53.155604 & -27.794268 & 125 & 8.40 & -0.45 & 0.62 & 0.56 \\
6568 & $\phn 1.76 _{-0.08}^{+0.11}$& 53.151289 & -27.792543 & 251 & 7.81 & -1.09 & 0.35 & 0.11 \\
6995 & $\phn 1.79 _{-0.06}^{+0.13}$& 53.154076 & -27.790955 & 232 & 9.95 & 0.41 & 1.39 & 1.33 \\
7063 & $\phn 1.76 _{-0.03}^{+0.07}$& 53.160417 & -27.790366 & 215 & 9.61 & 1.34 & 0.37 & 0.50 \\
7100 & $\phn 1.87 _{-0.08}^{+0.15}$& 53.154010 & -27.790844 & 236 & 10.00 & 1.55 & 1.49 & 2.13 \\
7133 & $\phn 1.81 _{-0.03}^{+0.09}$& 53.155016 & -27.790425 & 233 & 8.92 & 0.73 & 0.42 & 0.64 \\
7564 & $\phn 1.76 _{-0.15}^{+0.36}$& 53.164471 & -27.789714 & 279 & 7.26 & -1.39 & 0.30 & 0.16 \\
7621 & $\phn 1.79 _{-0.10}^{+0.35}$& 53.158685 & -27.789563 & 236 & 7.80 & -1.24 & 0.55 & 0.40 \\
8032 & $\phn 1.90 _{-0.08}^{+0.21}$& 53.161883 & -27.788222 & 290 & 8.31 & -0.17 & 0.52 & 0.59 \\
8064 & $\phn 1.93 _{-0.09}^{+0.27}$& 53.161802 & -27.788231 & 289 & 8.17 & -0.28 & 0.51 & 0.59 \\
\enddata
\tablenotetext{a}{Compact galaxy, XDF463.}
\tablenotetext{b}{Photometric redshift derived with EAZY (Brammer~{et al.}~\citeyear{brammer08}).}
\tablenotetext{c}{Projected distance from XDF463 (kpc).}
\tablenotetext{d}{Stellar mass ($M_\odot$), star formation rate ($M_\odot$yr$^{-1}$) and the rest-frame $UVJ$ colors (mag) derived by SED fitting with FAST (Kriek~{et al.}~\citeyear{kriek09}).}
\end{deluxetable*}

\bibliography{adssample}

\end{document}